# EFFECTS OF MACROION GEOMETRY AND CHARGE DISCRETIZATION IN CHARGE REVERSAL

Arup K. Mukherjee

Department of Physics
Chancellor College
University of Malawi
Box 280
Zomba, Malawi

# ABSTRUCT

The effects of discrete macroion surface charge distribution and valences of these surface charges and counterions on charge reversal have been studied for macroions of three different geometries and compared with those of continuous surface charge distributions. The geometry of the macroion has been observed to play an important role in overcharging in these cases. The interplay of valences of discrete microions and counterions have noticeable effects on overcharging efficiency. For some valence combinations overcharging the macroion with discrete surface charge is seen to be more efficient than those with continuous charge distribution. The calculations have been performed using a previously developed energy minimization simulation technique. For comparison purposes the corresponding continuous charge distribution cases together with an analytical model derived from modified Scatchard approach (for continuous charge distribution) have been evaluated.

INTRODUCTION

Overcharging or overscreening of a macroion, which is responsible for charge inversion, is a very interesting phenomenon which is often at odds with the classical mean field theories. This counterintuitive effect is thought to be a reason behind one of the mechanisms of like-charge attraction [1-4]. Due to its possible application to colloid chemistry and particularly to biological systems vis-a-vis numerous technological innovations [5] overcharging phenomena has attracted a lot of attentions [1-11]. One interesting question related to overcharging, which is often overlooked, is whether there is any effect of macroion geometry on it. This is especially important in biological systems where the macromolecules can have different geometries [1]. Only recently, a couple of papers have been reported comparative studies on such geometrical effects on overcharging where three different geometries of a macroion, with same surface area [1] or volume [12,13] have been considered.

It has been observed that the nature of the surface charge distribution of the macroion plays an important role in overcharging efficiency. Traditionally macroion surface charge is often considered as continuous or homogeneous. But the evidence and importance of discrete surface charge distribution has also been established experimentally [14-16]. For example, Litton and Olson [14] have studied the adsorption of sodium dodecyl sulfate on quartz surface and carbonyl latex. They found rather strong evidence of surface inhomogeneity on particle deposition rates. Kihira *et al* [15] experimentally measured coagulation rates in dispersion of spheres of sulfate lattices, silica and cerium oxide as a function of the ionic strength. They employed the results to interpret the discrepancies between theoretical prediction and experimental results of slow coagulation kinetics. It was found that the agreement between the experimentally measured and theoretically predicted stability ratios can only be established if the surface charges of the spheres are assumed to be heterogeneous. As another example, the aggregation rates of amidine and sulfate polystyrene latex particles were measured as a function of polymer dose and ionic strength by F. Bouyer *et al* [16]. They found that fast aggregation regime was increased with the increase of ionic strength. They suggested that this effect is characteristic for results from the existence of surface charge heterogeneities.

Indeed, one of the reasons behind surface charge inhomogeneity is the discrete surface charge of the macroion. Due to strong experimental evidences of its existence discrete nature of macroion surface charge has been of great interest recently [4, 5, 14-21].

In this paper, like our previous studies [1,12], the overcharging phenomena will be studied employing different geometries such as spherocylinder, sphere and oblate spheroid for the macroion with discrete surface charge distribution and will be compared with those of *continuous charge* (CC) distribution. Comparison of results reveals that the overcharging efficiency depends not only on the combinations of valances of the *discrete colloidal charge* (DCC) and counterions but also on the geometry of the macroion.

## MODELS AND METHODS

The charge $Q$ of an isolated macroion can be treated as discrete by assuming it to be comprised of N number of small identical charged microions (DCC microions), each having a charge $-Z_d e$ [see Table 1], which are distributed over the macroion surface so that the total charge of the macroion is $Q = -NZ_d e$. The distribution of the DCC microions over the macroion surface and then overcharging the macroion by counterions have been performed in the following two steps applying a previously developed [1] energy minimizing technique (EM) : (a) for a (nearly) uniform distribution of the DCC, the lowest (neutral) energy configuration has been achieved within about three millions moves per DCC. The centers of the DCC microions are then held fixed at their respective minimized positions on the macroion surface. (b) For overcharging, the counterions have been distributed over the DCC microions using the same EM technique again [following same procedures as described in ref. 1] and about two millions moves have been attempted by each counterion to reach the *minimal* configuration.

The sizes of the DCC microions and the counterions are considered identical but the charges are opposite. A clearer picture of the geometry involved can be had in Figure 1, which shows the neutral state DCC and counterion distributions over the macroions having spherocylindrical and oblate spheroid geometries. The dashed lines indicate the surfaces of the macroions over which the centers of the DCC microions are distributed. The centers of the counterions are on the surfaces indicated by solid lines. These surfaces (solid lines) are one diameter of a DCC (or a counterion) away from the macroion surfaces (dashed lines). In case of CC distribution the macroion diameters have been readjusted so that the counterion-macroion closest approach remains exactly the same as in the discrete case. The total energy in CC case is comprised of counterion-macroion and counterion-counterion interactions while in DCC case it is DCC-DCC, DCC-counterion and Counterion-counterion interactions. For non-spherical macroions the calculations involve a geometrical parameter τ, which is defined as $L = r_m \tau$ [1] where L is the length of the cylinder of a spherocylindrical macroion or the distance between the centers of curvature of the two hemispherical surfaces of an oblate spheroid. $r_m$ is the adjusted radius of the non-spherical macroion. The other relevant physical and model parameters are given in Table 1.

**TABLE I.** Physical parameters and symbols used in the present simulations.

--------------------------------------------------------------------------------

Room temperature T = 298 K

Relative permittivity $\varepsilon_r = 16.0$

Valence of counterions $Z_c = 1, 2$

Valence of DCC microions $Z_d = 1, 2$

Valence of macroion $Z_m = 180$

Counterion and DCC microion radius = 1.8 Å

Macroion (spherical)-counterion distance of closest approach = 28.76 Å

--------------------------------------------------------------------------------

---

From a proposed [1] theoretical model derived by modifying Scatchard approach, a relation between the energy of overcharging relative to neutral state and the number of overcharging counterions $n$ for CC distribution can be obtained as

$$\Delta E_n = \frac{\Delta E_1}{G} \frac{n}{N_c}[GN_c + 1 - n]$$

(1) Where $\Delta E_1$ is the energy of the first overcharge relative to the neutral state, G is an arbitrary fit parameter and $N_c = \frac{Z_m}{Z_c}$. Note that equation (1) does not contain any term related to the geometry of the macroion and thus can fit the simulation data for any geometry.

## RESULTS AND DISCUSSIONS

In this study we have focused more on non-spherical macroions than on spherical ones. The spherical macroions have been considered only for comparison purpose. One reason is that spherical cases have already been treated earlier in the literatures. For instance, a good discussion on spherical macroion cases can be found in ref.[4,17,18].

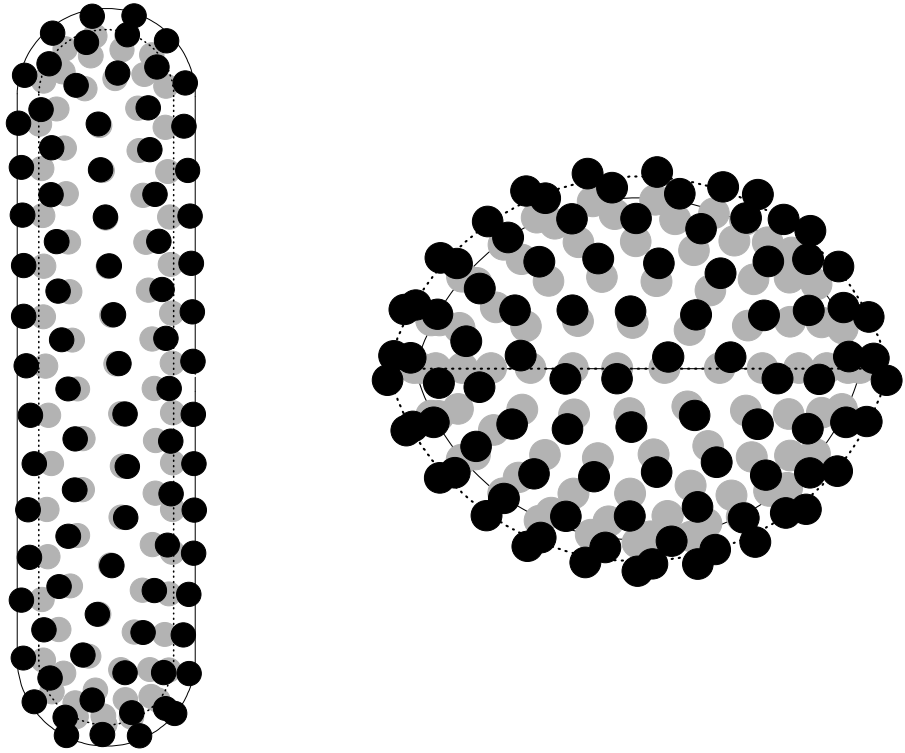

Fig. 1 The neutral state *minimal* distributions of monovalent counterions (deep black dots) and DCC microions (light black dots) over the surfaces of (A) a spherocylinder ($\tau = +10.0$) and (B) an oblatespheroid ($\tau = -0.5$). The surfaces of the two non-spherical macroions are equal to the surface area of a spherical macroion ($\tau = 0.0$) with radius R = 23.36 Å. There are 180 counterions distributed over equal number of monovalent DCC (only side views are shown). In both the cases complete ion-pairings are visible.

Figure 1 shows the *minimal* configurations (neutral states) of the counterions and the DCC microions for two different non-spherical geometries of the macroion. The neutral state means the global charge neutrality of a single macroion where the total charge of the macroion is equal to that of counterions. Both the counterion and the DCC are monovalent. Similar complete ion-pairings, as shown in figure 1, have been observed in all symmetric valance cases. In figure 2 the overcharging curves for three different geometries of the macroion is shown. For comparison both the cases (CC and DCC) with mono- and divalent counterions on the macroion surface (CC case) and on DCC microions (DCC case) have been drawn for each macroion geometry and the corresponding overcharging curves are specified according to Table 2. For convenience the combination of the valences of the discrete surface charge $Z_d$ and those of the counterion $Z_c$ will be indicated by [$Z_d : Z_c$]. Figure 2 clearly depicts that, in case of DCC, if the overcharging is performed with monovalent counterions [1:1, 2:1] the curves are very close to one another despite the valence of the DCC microions and also to that of CC. But if it is divalent [1:2, 2:2], the energy gain is much higher than that of monovalent overcharging for both CC and DCC cases which is a very reasonable well known aspect of overcharging and have been observed in many cases [1,4,12,13,18]. Another interesting point is that the gain is remarkably higher if the DCC microions are monovalent [1:2]. It implies that the total attraction between monovalent DCC and divalent counterions is much higher than the total repulsion among members of each species. Actually for [1:2] case the number of divalent counterions is almost half of the monovalent DCC microions and thus the average distance between the counterions is approximately twice as much the average distance between the DCC microions.

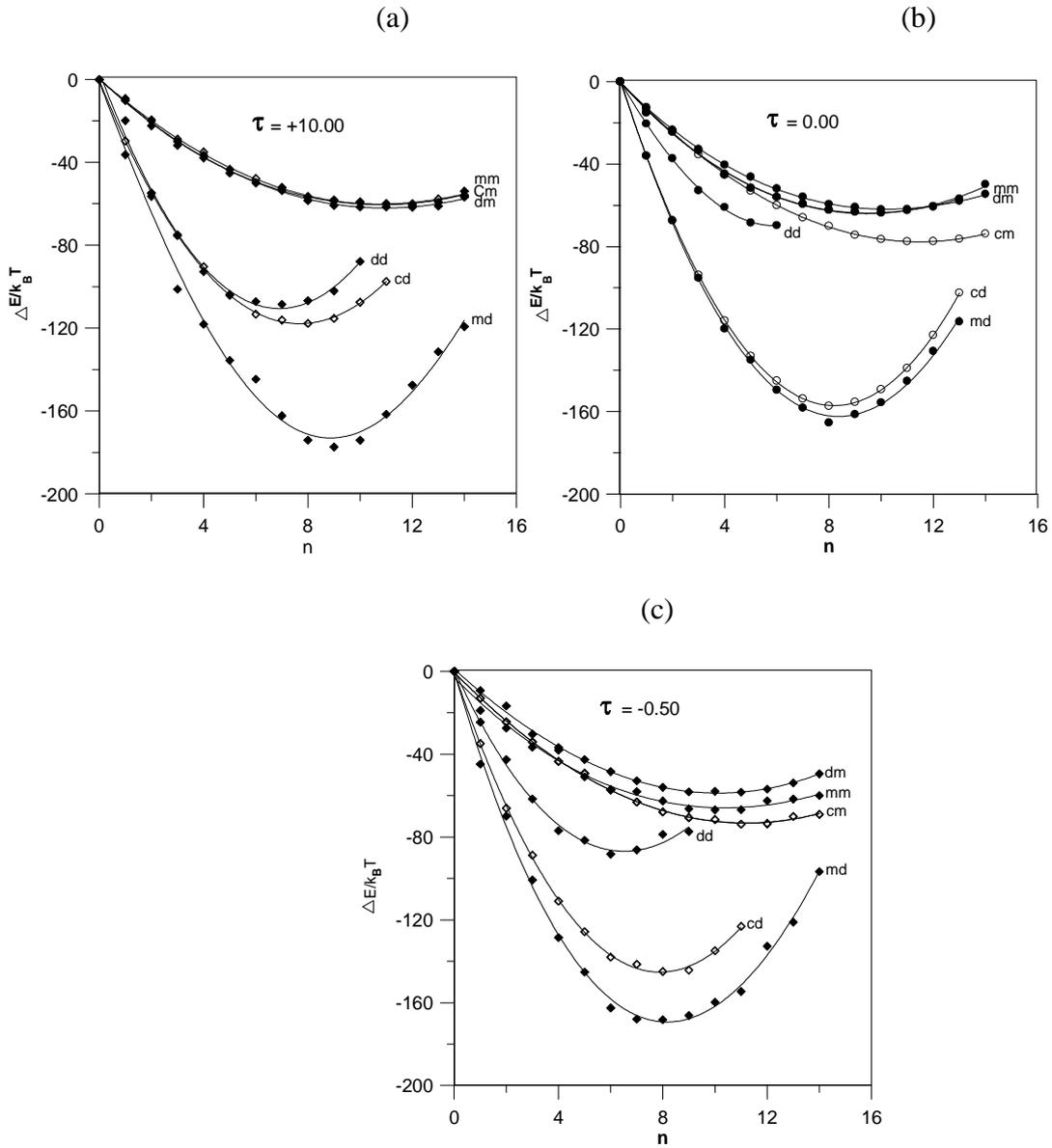

**Fig. 2** The overcharging curves for different combinations of valences of DCC and counterions for macroions with (a) $\tau = +10.0$, (b) $\tau = 0.0$ and (c) $\tau = -0.5$. The curves are identified by a couple of letters explained in Table 2. The first and the second letters indicate the valences of a DCC and of a counterion respectively. The monovalent is indicated by 'm' and the divalent is by 'd'. The CC cases (open symbols) are identified by a 'c' at the beginning and the next letter indicates the valence of the overcharging counterion. The solid lines for CC cases are obtained from equation 1 while other solid lines are polynomial fits to guide the eyes.

**Table 2.** Comparison between the neutral state energies (in units of $k_BT$) of different combinations of valences of DCC and counterions (CI) for both discrete and continuous cases for three different geometries: oblate spheroid ($\tau < 0$) sphere ($\tau = 0$) and spherocylender ($\tau > 0$). The valences of counterions in CC cases are indicated in the parentheses adjacent to their energies.

| Valences | | | $\tau = -0.5$ | | $\tau = 0.0$ | | $\tau = +10.0$ | |
|---|---|---|---|---|---|---|---|---|
| DCC | CI | Sym. | DCC | CC(val.) | DCC | CC(val.) | DCC | CC(val.) |
| 1 | 1 | mm | -832.4 | -21809.8(1) | -818.2 | -21364.6(1) | -863.4 | -21534.7(1) |
| 1 | 2 | md | -1428.8 | | -1402.0 | | -1434.5 | |
| 2 | 1 | dm | -1712.0 | -22516.3(2) | -1630.0 | -22036.2(2) | -1708.6 | -22232.9(2) |
| 2 | 2 | dd | -2838.7 | | -2827.8 | | -2859.3 | |

The divalent overcharging curves [cd] of CC cases are always in between [1:2] and [2:2] curves of DCC cases for all geometries, but for $\tau$ = -0.5 and 0.0, those are close to [1:2] DCC curves with significant energy gain specially for $\tau = 0.0$ while it is close to [2:2] curves for $\tau = +10$. This is due to the correlation energy which '*scales like* $-Z^{\frac{3}{2}}$ '[18] for $\tau = 0.0$ while it is $-Z$ for $\tau > 0$ [13]. The energy gains of [1:2] curves are approximately the same for all geometries while the gains of [2:2] curves vary with geometry. The overall situation suggests that overcharging by multivalent counterions is more efficient than that of monovalent for both CC and DCC cases, but the energy gain depends significantly on macroion geometry. The solid lines for CC cases (cm and cd) of all geometries are obtained from equation 1. The required fit parameters are given in Table 3.

**Table 3**. Fit parameters G of equation 1 for the CC cases in Fig. 2. The 3$^{rd}$ and 4$^{th}$ columns indicate the counterion valance and the corresponding curves respectively.

| $\tau$ | G | valence | symbol |
|---|---|---|---|
| -0.5 | 0.116 | 1 | cm |
|  | 0.165 | 2 | cd |
| 0.0 | 0.119 | 1 | cm |
|  | 0.171 | 2 | cd |
| +10.0 | 0.121 | 1 | cm |
|  | 0.157 | 2 | cd |

Although the overcharging curves in Figure 2, which are produced relative to the energy of neutral states of each individual DCC microion-counterion combinations are in relatively close proximity to one another, their positions in the graphs could be dramatically different if the total energies were considered. The neutral state energies of the curves of Figure 2 are tabulated in Table 2. Those energies of a specific combination of a specific case (CC or DCC) are comparable for all geometries but they are drastically different when compared between different combinations of valencies. On the other hand, the energy of CC overcharging is at least seven times lower than the corresponding energy of DCC overcharging. The deviation is due to the repulsion among the microions in DCC cases. The repulsive energies among DCC microions are listed in Table 4.

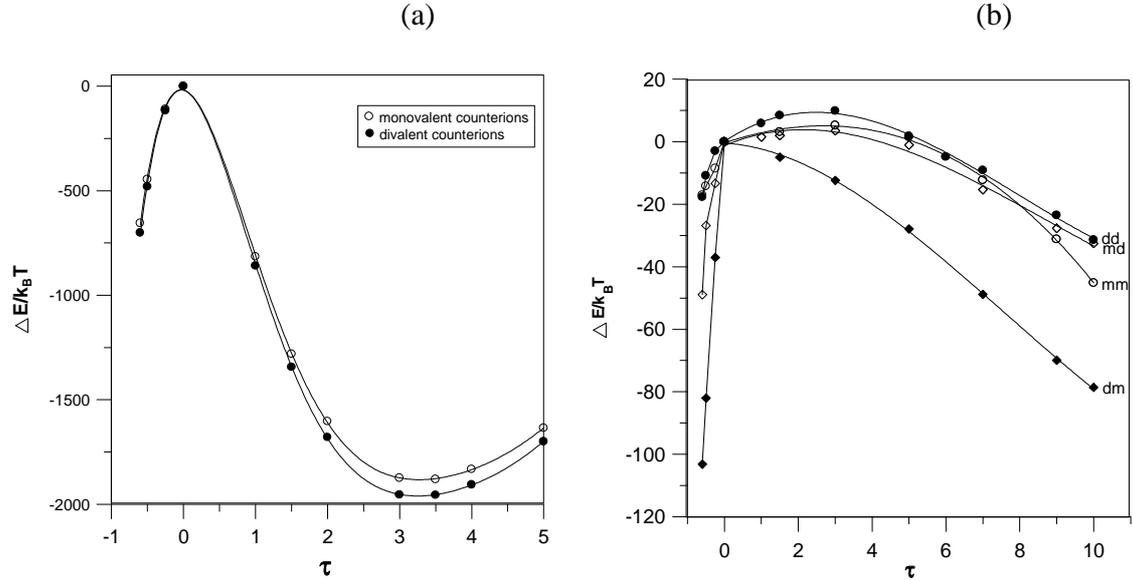

Fig. 3 Neutral state energies of a macroion as a function of τ. (a) CC case: the circles and the dots represent the monovalent and divalent neutralizing counterions respectively. (b) DCC case: the curves are identified using the same letters as in figure 3. The solid lines of (a) and the portion $0 < \tau < +10.0$ of (b) are polynomial fits to guide the eyes.

Figure 3 depicts the neutral state energy variation with macroion geometry and size (τ). For the CC case (Figure 3(a)) the neutral state energy varies smoothly with τ implying a rather continuous $\Delta E$ for $-0.6 < \tau < +10.0$ (see also [1,12]). In DCC case however (Figure 3(b)), it appears that there exists two different functions: (a) $-0.6 < \tau < 0.0$ and (b) $0.0 < \tau < +10.0$ for the two non-spherical geometries. A remarkable difference between the DCC and CC cases is that, while in case of CC, at some certain length L of the spherocylinder ($L = r_m \tau$ [1,12]) the neutral state energy shows a minimum (similar to overcharging), this feature is absent in case of DCC. The calculations have been performed up to $\tau = +20$ and no such minimum was found in case of DCC.

**Table 4.** The repulsive energies among DCC macroions in units of ($k_B T$).

| DCC Valence | $\tau = -0.5$ | $\tau = 0.0$ | $\tau = +10.0$ |
|---|---|---|---|
| 1 | 9881.917 | 19942.304 | 17806.604 |
| 2 | 20687.881 | 20709.442 | 18586.400 |

## CONCLUDING REMARKS

In this paper, using the previously developed energy minimization technique (EM), a comparative study of overcharging phenomena for macroions of different geometries has been done for both discrete and continuous cases. Comparison between different geometries of macroions reveals a generalized picture of the overcharging process, which throws lights on the question whether overcharging in DCC cases are stronger than CC cases. If the macroion charge is represented by monovalent DCC, overcharging the macroions of any geometry and with relatively higher internal charge by divalent counterions dominates over the CC case under identical conditions. The present study produces similar results for most of the features of overcharging a spherical macroion using MD technique reported earlier in literatures [4,17,18], for example, stronger overcharging in CC case than in DCC for [1:1] and [2:2] for all geometries. But in case of 1:2 the overcharging is weaker than in CC case, which is practically opposite to the results of ref.[18] for $Z_m = 180$. However, this pair of curves (at this macroion charge) is very close to each other for spherical macroion in both the studies, predicting very similar physical features within the range of statistical uncertainty. Nevertheless, comparison among the results of the three geometries indicates that the DCC overcharging [1:2] is stronger than CC in all cases. This has also been observed in lower macroion charges. Note that, all results using the EM technique are supposed to be deviated [1,12] with the increase of overcharging counterions (n) from those of ref. [4,17,18] due to the difference in the type of potentials used for colloid-counterion interactions. Hard sphere potential has been used in this study, while, in ref.[4,17,18] the excluded volume interaction has been taken into account with repulsive Lennard-Jones potential which allows counterions to penetrate into the (soft) macroion.

For lower macroion charge, for example $Z_m = 60$, the overcharging curve (not shown here) for [1:1] in DCC case has been observed to lie above the curve for monovalent overcharging in CC case. But it is opposite for [1:2] DCC when compared with divalent overcharging in CC case. This remarkable feature has also been reported in ref.[17].

It is important to note that the equation 1 is in good agreement with the simulation results for CC cases which implies that the proposed theoretical model can explain the overcharging curves for all counterion valences and macroion geometries. In future, a more elaborated study on different macroion charges and sizes will be presented for further generalization of the overcharging phenomenon for a single and also for a couple of interacting macroions for discrete colloidal charge cases and compared with those for conventional continuous charge cases.